\documentclass[12pt]{article} %***
\usepackage[sectionbib]{natbib}
\usepackage{array,epsfig,fancyhdr,rotating, xcolor}
\usepackage[]{hyperref}  %<----modified by Ivan
\usepackage{sectsty, secdot}
\sectionfont{\fontsize{12}{14pt plus.8pt minus .6pt}\selectfont}
\renewcommand{\theequation}{\thesection\arabic{equation}}
\subsectionfont{\fontsize{12}{14pt plus.8pt minus .6pt}\selectfont}
\usepackage{amsmath}
\usepackage{amssymb}
\usepackage{amsfonts}
\usepackage{multirow}
\usepackage{amsthm}

\newtheorem{theorem}{Theorem}
\newtheorem{lemma}{Lemma}

\newtheorem{proposition}{Proposition}
\theoremstyle{definition}

\newtheorem{remark}{Remark}
\DeclareMathOperator*{\argmax}{argmax}

\newcommand{\ve}[1]{{\mbox{\boldmath ${#1}$}}}

\begin{document}

%%%%%%%%%%%%%%%%%%%%%%%%%%%%%%%%%%%%%%%%%%%%%%%%%%%%%%%%%%%%%%%%%%%%%%%%%%%%%%%%%%%%%%%%%%%%%%%%%%%%%%%%%%%%%%%%%%%%%%%%%%%%
%%%%%%%%%%%%%%%%%%%%%%%%%%%%%%%%%%%%%%%%%%%%%%%%%%%%%%%%%%%%%%%%%%%%%%%%%%%%%%%%%%%%%%%%%%%%%%%%%%%%%%%%%%%%%%%%%%%%%%%%%%%%

\renewcommand{\baselinestretch}{2}

\markright{ \hbox{\footnotesize\rm Statistica Sinica
%{\footnotesize\bf 24} (201?), 000-000
}\hfill\\[-13pt]
\hbox{\footnotesize\rm
%\href{http://dx.doi.org/10.5705/ss.20??.???}{doi:http://dx.doi.org/10.5705/ss.20??.???}
}\hfill }

\markboth{\hfill{\footnotesize\rm Yue Wang} \hfill}
{\hfill {\footnotesize\rm RAPLS} \hfill}

\renewcommand{\thefootnote}{}
$\ $\par

%%%%%%%%%%%%%%%%%%%%%%%%%%%%%%%%%%%%%%%%%%%%%%%%%%%%%%%%%%%%%%%%%%%%%%%%%%%%%%%%%%%%%%%%%%%%%%%%%%%%%%%%%%%%%%%%%%%%%%%%%%%%

\fontsize{12}{14pt plus.8pt minus .6pt}\selectfont \vspace{0.8pc}
\centerline{\large\bf Residual-based Alternative  Partial Least Squares}
\vspace{2pt} 
\centerline{\large\bf for Generalized Functional Linear Models}
\vspace{.4cm} 
\centerline{Yue Wang$^1$, Xiao Wang$^2$, Joseph G. Ibrahim$^3$, and Hongtu Zhu$^3$} 
\vspace{.4cm} 
\centerline{\it $~^1$Department of Biostatistics and Informatics, Colorado School of Public Health}
\centerline{\it $~^2$Department of Statistics, Purdue University}
\centerline{\it $~^3$Department of Biostatistics, University of North Carolina at Chapel Hill}
 \vspace{.55cm} \fontsize{9}{11.5pt plus.8pt minus.6pt}\selectfont

%%%%%%%%%%%%%%%%%%%%%%%%%%%%%%%%%%%%%%%%%%%%%%%%%%%%%%%%%%%%%%%%%%%%%%%%%%%%%%%%%%%%%%%%%%%%%%%%%%%%%%%%%%%%%%%%%%%%%%%%%%%%

\begin{quotation}
\noindent {\it Abstract:}
Many biomedical studies collect high-dimensional medical imaging data to identify biomarkers for the detection, diagnosis, and treatment of human diseases. Consequently, it is crucial to develop accurate models that can predict a wide range of clinical outcomes (both discrete and continuous) based on imaging data. By treating imaging predictors as functional data, we propose a residual-based alternative partial least squares (RAPLS) model for a broad class of generalized functional linear models that incorporate both functional and scalar covariates. Our RAPLS method extends the alternative partial least squares (APLS) algorithm iteratively to accommodate additional scalar covariates and non-continuous outcomes. We establish the convergence rate of the RAPLS estimator for the unknown slope function and, with an additional calibration step, we prove the asymptotic normality and efficiency of the calibrated RAPLS estimator for the scalar parameters. The effectiveness of the RAPLS algorithm is demonstrated through multiple simulation studies and an application predicting Alzheimer's disease progression using neuroimaging data from the Alzheimer’s Disease Neuroimaging Initiative (ADNI).

\vspace{9pt}
\par
\end{quotation}\par

\def\thefigure{\arabic{figure}}
\def\thetable{\arabic{table}}

\renewcommand{\theequation}{\thesection.\arabic{equation}}

\fontsize{12}{14pt plus.8pt minus .6pt}\selectfont
%-------------------------------------------------------

\section{Introduction}

An important class of problems in medical imaging research is to identify imaging patterns associated with clinical outcomes of interest.   
   Suppose that we observe a sample of $n$ {\it i.i.d} subjects 
   $\{y_i, {x}_{i}(\cdot),  {\bf z}_i\}_{i=1}^n$ from the joint distribution of $(y, {x}(\cdot),  {\bf z})$,
 where
${\bf z}\in \mathbb R^q$ represents $q$-dimensional covariates, $y\in \mathbb R$ is a continuous or discrete outcome variable of interest, and $x(\cdot)$
%${\bf x}_{i}=(x_{i}(s):  s\in \mathcal S)$
represents the functional imaging data, observed 
at a set of grid points in a nondegenerate and compact space $\mathcal S\subset \mathbb{R}^K$ for some positive integer $K$.  
%for the $i$-th subject, where $K>0$ is a positive integer.  
% Without loss of generality, 
% we assume that $\mathbb{E}\{x(s)\}=0$ for all $s \in \mathcal{S}$ and $\mathbb{E}({\bf z}) = {\bf 0}$. 
% We also assume that all imaging/functional data %are observed without measurement errors
% have been smoothed over $\mathcal{S}$.
%Treating the imaging predictors as functional data, researchers often use
Functional regression models are useful tools to examine how these imaging predictors impact the outcome while adjusting for scalar covariates. 
An important class of functional regression models is the generalized functional linear model (GFLM, \cite{muller2005generalized}).
%The generalized functional linear model (GFLM, \cite{muller2005generalized}), is a popular method to examine associations between scalar/functional covariates and an outcome of interest.
Specifically, for some underlying true parameters $\alpha_0^*$, $\ve \alpha^*$ and $b^*(\cdot)$, let $\eta_i^* = \alpha_0^* + {\bf z}_i^\intercal \ve \alpha^* + \int_{\mathcal{S}} x_i(s)b^*(s)ds$ denote the true linear predictor for the $i$-th subject. Given $\eta_i^*$, GFLM specifies the distribution of $y_i$ within the exponential family; that is, 
\begin{equation}\label{rev:1}
    p(y_i, \eta_i^*) = h(y_i)\exp\{\eta_i^* T(y_i) - A(\eta_i^*)\}
\end{equation}
for some $h(\cdot), T(\cdot)$ and $A(\cdot)$.  
%Any model that can be written in the form of \eqref{rev:1} is called a generalized functional linear model (GFLM, \cite{muller2005generalized}). 
A special case of the GFLM \eqref{rev:1} is the partially functional linear model (PFLM) where $y_i = \alpha_0^* + {\bf z}_i^\intercal \ve \alpha^* + \int_{\mathcal{S}} x_i(s)b^*(s)ds + \epsilon_i$ with $\epsilon_i \sim N(0, \sigma^2)$.
%with both scalar and functional covariates, called the partially functional linear model (PFLM) hereafter.
When no scalar covariates ${\bf z}_i$ exist,  PFLM reduces to the simple functional linear model (SFLM), which have been extensively studied in the literature; see \cite{Ramsay2005,Ferraty2006,horvath2012inference,Morris2015,Wang2015, Kong2016} and the references therein. 
Extensions of (\ref{rev:1}) for predicting survival and longitudinal outcomes 
have also been developed in the literature \citep{gellar2015, Lee2015, qu2016, li2017functional, wang2020partial}.

Estimating the functional parameter $b^*(\cdot)$ in \eqref{rev:1} requires dimensionality reduction due to the infinite dimension of $b^*(\cdot)$. The general approach is to approximate both $x_i(\cdot)$ and $b^*(\cdot)$ using a set of orthonormal basis functions, reducing the estimation of $b^*(\cdot)$ to estimating the coefficients associated with each basis function. Functional principal component analysis (FPCA) is a key technique for constructing data-driven basis functions; for an extensive review of FPCA, see \cite{Besse1986,RamsayDalzell1991,Boente2000335,James2000} and \cite{Hall2006}. The top principal components (PCs), which correspond to the largest eigenvalues and explain the most data variation, are then used to simplify GFLM into the classical generalized linear model (GLM). 
{
%\color{blue}
However, using principal components (PCs) in regression problems has two notable limitations. First, functional principal component analysis (FPCA) does not incorporate information from the outcome, meaning that the top PCs may not capture the relationship between the functional predictor and the outcome. This can result in suboptimal prediction or estimation accuracy \citep{cook2007fisher}. Second, when tail PCs are important for improving prediction, FPCA requires a large sample size to accurately estimate these PCs, which are associated with smaller eigenvalues \citep{Jung2009}. This poses a challenge in many medical imaging studies, where sample sizes are often constrained due to budgetary limitations.
}
% empirical eigenfunctions may {\it not} be accurate when the sample size is limited \citep{Jung2009}, 
% %accurately estimate the corresponding true eigenfunctions, 
% especially for tail eigenfunctions corresponding to small eigenvalues. %Indeed, in real neuroimaging applications with limited sample sizes, when the underlying $b(\cdot)$ is (partly) aligned with tail eigenfunctions, FPCA may fail to recover $b(\cdot)$, leading to inaccurate prediction.
% On the other hand, the reproducing kernel Hilbert space (RKHS) approach \citep{yuan2010, CaiYuan2012} provides an attractive alternative by assuming the unknown $b^*(\cdot)$ resides in an RKHS embedded with a pre-specified kernel. However, when the reproducing kernel is mis-specified, the RKHS approach may have poor performance. 

To address the limitations of FPCA, partial least squares (PLS) methods have been developed for functional regression models. Unlike FPCA, PLS incorporates information from both the covariates and the outcome when constructing basis functions, eliminating the need to compute empirical eigenfunctions. Early PLS methods were designed for simple functional linear models (SFLMs, \citealp{Preda2005b}), using an iterative procedure to estimate PLS components by maximizing the covariance between the outcome and a linear form of the functional predictor. Later, \cite{Hall2011} introduced an alternative approach, known as alternative partial least squares (APLS), which provides a different set of basis functions that span the same space as the PLS basis but are computationally easier to obtain. 
Since then, theoretical advancements have also been made in functional PLS (FPLS) for SFLMs \citep{Preda2005b,Preda2005, Preda2007, escabias2007functional, reiss2007functional, kramer2008penalized, aguilera2010using, Hall2011, aguilera2016penalized, febrero2017functional}.
However, a key limitation of these PLS methods is their inherent design for linear relationships, making them less adaptable to nonlinear models. While extensions of PLS for nonlinear models exist, most involve repeatedly fitting GLMs, obtaining residuals, and using these residuals to construct PLS components \citep{bastien2005pls}. Consequently, in nonlinear models, PLS components no longer maximize the covariance between the outcome and covariates, which undermines one of the central advantages of PLS in linear settings and leads to poorer estimation and prediction. 

%------------------------------------------------
% Start from here; 
We propose a functional partial least squares (FPLS) method, called residual-based alternative partial least squares (RAPLS), for estimating the generalized functional linear model (GFLM) in (\ref{rev:1}), which fundamentally differs from existing nonlinear PLS methods. RAPLS extends the alternative partial least squares (APLS) procedure from \cite{Hall2011} to nonlinear models by iteratively fitting reweighted functional linear models. 
This approach adapts the iteratively reweighted least squares (IRLS) method \citep{green1984} to functional regression models and leverages the computational efficiency of the APLS approach. 
We also realize the gap that the existing theory for PLS has primarily been developed in linear settings, but theoretical justifications for nonlinear FPLS procedures are scarce. 
We bridge this gap by establishing the theoretical properties of the proposed RAPLS algorithm. Specifically, we establish the convergence rate of the RAPLS estimate for $b^*(\cdot)$, forming the foundation for proving the consistency of RAPLS estimates. 
We allow the number of components used in the RAPLS algorithm to diverge with the sample size, reflecting the infinite-dimensional nature of functional data. 
%which necessitates an infinite number of FPLS components.
Moreover, we develop a calibrated estimator for the scalar covariate that is asymptotically normal and efficient. Finally, we compare the finite-sample performance of the RAPLS algorithm with multiple existing methods using simulated data sets and an application focused on predicting the progression of Alzheimer's Disease (AD).

Throughout the paper, for any vector ${\bf v} \in \mathbb{R}^d$, we use $v_j$ to denote the $j$-{th} element of $\bf v$ for $j = 1, \ldots, d$. For any matrix ${\bf M} \in \mathbb{R}^{n \times d}$, let $ {\bf m}_j$ and $m_{ij}$ denote the  $j$-{th} column and $(i,j)$-th entry of $\bf M$,  respectively for $i = 1, \ldots, n$ and $j = 1, \ldots, d$. 
{For ease of notation, we use a single notation $\|\cdot\|$ to denote the $\ell_2$-norm for vectors, matrices, functions, operators, and kernels.} Specifically, let $\|{\bf v}\|^2 = \sum_{j=1}^d v_j^2$ and $\|{\bf M}\| = \sup_{\|{\bf v}\| = 1}\|{\bf Mv}\|$.  
For any square-integrable function $f(\cdot)$, we let $\|f\|^2 = \int f^2(s)ds$. For any positive semi-definite kernel function $K(\cdot, \cdot)$, we let 
%$K(f)(t) = \int K(s,t)f(s)ds$, 
$\|K\| = \sup_{\|{f}\| = 1}\|{K(f)}\|$. Also, we denote $\|K\|_F^2 = \iint K(s,t)^2 dsdt$.  We use ${\bf I}_d$ and ${\bf 1}_d$ to denote the $d\times d$ identity matrix and the $d$-dimensional vector with all ones, respectively.

%-------------------------------------------------------
\section{Population-level RAPLS for GFLMs}\label{pop:RAPLS}

In this section, we introduce the population-level RAPLS for GFLM (\ref{rev:1}). Recall that the true linear predictor is $\eta^* = \alpha_0^* + {\bf z}^\intercal \ve \alpha^* + \int_{\mathcal{S}} x(s)b^*(s)ds$. 
Without loss of generality, 
we assume that $\mathbb{E}\{x(s)\}=0$ for all $s \in \mathcal{S}$ and $\mathbb{E}({\bf z}) = {\bf 0}$.
For any random variable $w \in \mathbb{R}$, define
$
m({\bf z}, w) = w
- {\bf z}^\intercal \{\mathbb{E}({\bf z}^{\otimes 2})\}^{-1} \mathbb{E}({\bf z} w)$,
where ${\bf z}^{\otimes 2} = {\bf zz}^{\intercal}$. Since $\mathbb{E}({\bf z}) = {\bf 0}$, applying the function $m({\bf z}, \cdot)$ to both sides of $\eta^*$ yields
\begin{equation}\label{012722:1}
    m({\bf z}, \eta^*) = \int_{\mathcal{S}} m\left({\bf z}, x(s) \right) b^*(s) ds. 
\end{equation}
One can view $m({\bf z}, \eta^*)$ and $m\left({\bf z}, x(s) \right)$ as the {\it residual} of $\eta^*$ and $x(s)$, respectively, after removing the effect of $\bf z$. We also define the covariance kernel of the residual process $\left\{ m({\bf z}, x(s)) \right\}_{s \in \mathcal{S}}$ by $\mathcal{C}(s,t) = \mbox{cov}\left( m({\bf z}, x(s)), m({\bf z}, x(t)) \right)$ for any $s, t \in \mathcal{S}$. Note that $\mathcal{C}$ does not depend on any specific realization of ${\bf z}$.

Consider first an oracular scenario where the true linear predictor $\eta^*$ is known. 
According to \cite{Preda2005}, 
the FPLS basis functions for (\ref{012722:1}), denoted by $\{\rho_k(\cdot)\}_{k \geq 1}$, can be obtained by sequentially maximizing
\begin{align}\label{RAPLS5.1}
\mbox{cov}\left(m({\bf z}, \eta^*) - g_{k-1}\left(m({\bf z}, x)\right), \int_{\mathcal{S}} m\left({\bf z}, x(s) \right)\rho_k(s)ds\right)
\end{align}
subject to 
\begin{equation*}
\iint_{\mathcal{S}^2}\rho_j(s)\mathcal{C}(s,t)\rho_k(t)dsdt = 0~~~\mbox{for}~~ 1 \leq j \leq k-1~~~\mbox{and}~~~ \iint_{\mathcal{S}^2} \rho_k(s) \mathcal{C}(s,t) \rho_k(t) dsdt = 1, \end{equation*}
%where $\mathcal{C}(s,t) = \mbox{cov}(m({\bf z}, x(s)), m({\bf z}, x(t)))$, 
where 
$g_k(m({\bf z}, x)) =  \int_{\mathcal{S}} m({\bf z}, x(s))\tilde{b}^*_k(s)ds$, and $\tilde{b}^*_k(s)$ is the orthogonal projection of $b^*(s)$ onto the space spanned by $\rho_1(\cdot), \ldots, \rho_k(\cdot)$ under the $\ell_2$-norm.
%-----------------------------------------------

{
%\color{blue}
According to \eqref{RAPLS5.1}, the first FPLS basis function $\rho_1(\cdot)$ can be obtained by maximizing 
\(
%\mbox{cov}\left( m({\bf z}, \eta^*), \int_{\mathcal{S}} m({\bf z}, x(s)) \rho_1(s) ds \right) = 
\iint_{\mathcal{S}^2} \rho_1(s) \mathcal{C}(s,t) b^*(t) ds dt   
\)
subject to $\iint_{\mathcal{S}^2} \rho_1(s) \mathcal{C}(s,t) \rho_1(t) dsdt = 1$. 
Some functional calculus yields that  
$\rho_1(t)$ is proportional to $\mathcal{C}(b^*)(t)$, where $\mathcal{C}$ is an operator that maps $b(t)$ to $\int_{\mathcal{S}} \mathcal{C}(s,t)b(t)dt$. 
% A limitation of the FPLS basis functions in  (\ref{RAPLS5.1}) lies in
% their iterative nature,
% introducing both theoretical and computational complexities. 
Indeed, this result can be extended to all FPLS basis functions obtained by maximizing \eqref{RAPLS5.1}. 
}
% Thus, we extend the APLS method in \cite{Hall2011} to \eqref{012722:1}
% %develop the alternative partial least squares (APLS) for $\{\rho_k(\cdot)\}_{k \geq 1}$ 
% to address this limitation.
\begin{lemma}\label{lemma1}
Given $\{\rho_i(\cdot)\}_{i=1}^j$,  $\rho_{j+1}(\cdot)$ is unique up to sign change and  determined by
\begin{align*}
\rho_{j+1}(s) = c_0\left[\mathcal{C}\left(b^*(s) - \sum_{l=1}^{j}\left\{\int_{\mathcal{S}} b^*(t)\rho_{l}(t)dt \right\}\rho_{l}(s)\right) + \sum_{l=1}^{j} c_l \rho_l(s)\right], 
\end{align*}
where 
{
%\color{blue}
$\{\rho_j(\cdot)\}_{j \geq 1}$ are defined in (\ref{RAPLS5.1}),
}
$c_0$ is  uniquely defined up to a sign change, and
  $c_l$ are obtained by solving the linear system of $j$ equations
\begin{align*}
\iint_{\mathcal{S}^2} \rho_{l}(s) \rho_{j+1}(s) \mathcal{C}(s,t)dsdt = 0, \mbox~~ l=1, \ldots, j.
\end{align*}
\end{lemma}
The proof is similar to that of Theorem 3.1 in \cite{Hall2011}, which is omitted. 
An immediate observation is that
the space spanned by $\rho_{1}(s),\cdots,\rho_p(s)$ is the same as that spanned by $\mathcal{C}(b^*),\cdots,\mathcal{C}^p(b^*)$, where 
for $k > 1$, $\mathcal{C}^k(b^*) = \int_{\mathcal{S}} \mathcal{C}^{k-1}(b^*)(t)\mathcal{C}(s,t)dt$. 
{We call $\mathcal{C}^j(b^*)$ the $j$-th population-level residual-based APLS (RAPLS) basis function for  GFLM (\ref{rev:1}) for $j \geq 1$.}

{The next lemma, similar to Theorem 3.2 in \cite{Hall2011}}, gives conditions under which any square-integrable function $b(\cdot)$ can be written in a linear form of the RAPLS basis functions $\{\mathcal{C}^j(b)\}_{j=1}^\infty$.
\begin{lemma}\label{lemma2}
If  $\mathcal{C}(s,t)$ is positive definite, then any square-integrable function $b(\cdot)$ can be written as 
$b = \sum_{j=1}^\infty \gamma_j \mathcal{C}^j(b)$ for some constants $\{\gamma_j\}_{j=1}^\infty$, which converges in terms of the $\ell_2$-norm.
\end{lemma}
{
%\color{blue}
Letting $K(s,t) = \mbox{cov}(x(s), x(t))$, 
one can verify that $\mathcal{C}(s,t) = K(s,t) - \mathbb{E}\left\{{\bf z}^\intercal x(s)\right\} \left\{\mathbb{E} ({\bf z}^{\otimes 2})\right\}^{-1} \mathbb{E}\left\{{\bf z} x(t)\right\}$. 
}
Thus, 
the positive definiteness of $\mathcal{C}(s,t)$ can be guaranteed if $K(s,t)$ is positive definite, and $x(s)$ and $\bf z$ are not perfectly collinear, i.e., there do not exist $\nu_1(s), \ldots, \nu_{q_z}(s)$ such that $x(s) = \sum_{l=1}^{q_z} z_l \nu_l(s)$ for any $s \in \mathcal{S}$. 
%The proof is similar to that of Theorem 3.2 in \cite{Hall2011}, which is omitted here. 
Lemma \ref{lemma2} indicates that
in practice, we can truncate $\sum_{j=1}^\infty \gamma_j \mathcal{C}^j(b^*)$ to a finite number of components to approximate $b^*$. 
More specifically, for an arbitrary positive integer $p$, we can define $b_p^* = \sum_{j=1}^p \gamma_j^* \mathcal{C}^j(b^*)$ as the optimal $p$-th order RAPLS approximation of $b^*$. 
Thus, Lemma \ref{lemma2} guarantees that $\|b - b_p^*\| \rightarrow 0$ as $p \rightarrow \infty$, if $\mathcal{C}(s,t)$ is positive definite. This serves as the basis for the theoretical analyses in Section \ref{sec:conv:rate}. 
The ``optimal" coefficients $\gamma_1^*,\cdots,\gamma_p^*$ minimize  
\begin{align*}
\omega_p(\gamma_1,\ldots,\gamma_p) = &~ \mathbb{E}\bigg \{ m({\bf z}, \eta^*) - \sum_{j=1}^p \gamma_j \int_{\mathcal{S}} m({\bf z}, x(s)) \mathcal{C}^j(b^*)(s)ds\bigg \}^2. % \nonumber \\
%= &~ \mathbb{E}\big \{ \int_{\mathcal{S}} m({\bf z}, x(s)) b(s) ds - \sum_{j=1}^p \gamma_j \int_{\mathcal{S}} \mathcal{Z} (x(s)) (\mathcal{C})^j(s)ds\big\}^2.
\end{align*}
Letting $\ve \gamma^* = (\gamma_1^*, \ldots, \gamma_p^*)^\intercal$, we obtain $\ve \gamma^* = {{\bf H}^{*}}^{-1} \ve \beta^*$,
where ${\bf H}^* = (h_{jk}^*)_{j,k = 1, \ldots, p}$ and $\ve \beta^* = (\beta_j^*)_{j = 1, \ldots, p}$
with
\begin{equation}\label{hbeta:true}
h_{jk}^* =  \int_{\mathcal{S}} \mathcal{C}^{j+1}(b^*)(s)\mathcal{C}^{k}(b^*)(s)ds ~~\mbox{ and }~~
\beta_j^* =  \int_{\mathcal{S}} \mathcal{C}(b^*)(s)\mathcal{C}^{j}(b^*)(s)ds.
\end{equation}

%-------------------------------------------------------------------------------------

\section{Empirical RAPLS for GFLMs}\label{sec:3}

We start with the PFLM  $y_i = {\bf z}_i^\intercal \ve \alpha^* + \int_{\mathcal{S}} x_i(s)b^*(s)ds + \epsilon_i$ to gain some intuitions; here, we assumed $\mathbb{E}(y_i) = 0$ such that $\alpha_0 = 0$. 
The population RAPLS algorithm in Section \ref{pop:RAPLS} depends on two critical quantities: the residual covariance kernel $\mathcal{C}(s,t)$ and the true linear predictor $\eta^*$.
Since $\mathbb{E}({\bf z}_i) = {\bf 0}$ and $\mathbb{E}\{ x_i(s)\} = 0$ for all $s \in \mathcal{S}$, one can easily see that $\mathbb{E}\{m({\bf z}_i, x_i(s))\} = 0$ for all $s \in \mathcal{S}$. Thus, $\mathcal{C}(s,t)$ can be naturally estimated by its empirical counterpart 
\(
\widehat{\mathcal{C}}(s,t) = n^{-1} \sum_{i=1}^n m({\bf z}_i, x_i(s))m({\bf z}_i, x_i(t)).
\)
While the true linear predictor $\eta^*$ remains unknown in practice, it can be approximated by leveraging the relationship between the outcome $y$ and $\eta^*$. 
For PFLM, since $\mathbb{E}\left({y} \mid \eta^* \right) = \eta^*$, it is natural to use the outcome $y$ to approximate $\eta^*$. 
In particular, since
\begin{align}\label{est:basis1}
\mathcal{C}(b^*)(s) = \int_{\mathcal{S}} \mathcal{C}(s,t)b^*(t)dt = \mathbb{E} \left\{m({\bf z}, x(s)) m({\bf z}, y)\right\},
\end{align}
a natural estimate of $\mathcal{C}(b^*)(s)$ is
%\begin{equation*}
\(
\widehat{\mathcal{C}}(b)(s) = n^{-1} {\bf X}(s)^\intercal {\bf M}_Z {\bf y},
\)
%\end{equation*}
where ${\bf y} = (y_1, \ldots, y_n)^\intercal$, ${\bf X}(s) = \left( x_1(s), \ldots, x_n(s) \right)^\intercal$, 
${\bf M}_Z = {\bf I}_n - {\bf Z}({\bf Z}^\intercal {\bf Z})^{-1}{\bf Z}^\intercal$ with ${\bf Z} = ({\bf z}_1, \ldots, {\bf z}_n)^\intercal$.
Subsequently, for $j \geq 1$, we have
\(
\widehat{\mathcal{C}}^{j+1}(b) = \int_{\mathcal{S}} \widehat{\mathcal{C}}^j(b)(t) \widehat{\mathcal{C}}(s,t) dt.
\)
With these estimated basis functions, we  estimate $h_{jk}^*$ and $\beta_j^*$, respectively, with 
\[
\widehat{h}_{jk} = \int_{\mathcal{S}} \widehat{\mathcal{C}}^{j+1}(b)(s)\widehat{\mathcal{C}}^k(b)(s)ds~~\mbox{and}~~ \widehat{\beta}_j = \int_{\mathcal{S}}\widehat{\mathcal{C}}(b)(s)\widehat{\mathcal{C}}^{j+1}(b)(s)ds.
\]
Then, we calculate $\widehat{\ve \gamma} = \widehat{\bf H}^{-1}\widehat{\ve \beta}$, where $\widehat{\bf H} = (\widehat{h}_{jk})$ and $\widehat{\ve \beta} = (\widehat{\beta}_j)$. This leads to the RAPLS estimate of $b_{p}(s)$: 
\begin{equation}\label{est:pflm}
\widehat{b}_{p}^*(s) = \sum_{j=1}^p \widehat{\gamma}_j \widehat{\mathcal{C}}^{j}(b)(s), %Correspondingly, we have 
\end{equation}
where $\widehat{\gamma}_j$ is the $j$-th entry of $\widehat{\ve \gamma}$. Given $\widehat{b}_{p}(s)$, we obtain a plug-in estimator of $\ve \alpha^*$: \begin{equation}\label{alphaest:pflm}
\widehat{\ve \alpha}_p = ({\bf Z}^{\intercal}{\bf Z})^{-1} {\bf Z}^\intercal\left({\bf y} - \int_{\mathcal{S}} {\bf X}(s) \widehat{b}_{p}(s)  ds\right).
\end{equation}

%-------------------------------------------------

Next, we extend this RAPLS procedure to nonlinear GFLMs where
$\mathbb{E}\left( y \mid \eta^* \right) \neq \eta^*$. 
Our idea is to iteratively approximate the model (\ref{rev:1}) with a sequence of PFLMs. 
Let $\alpha_0^{(m)}$, $\ve \alpha^{(m)}$, and $b^{(m)}(\cdot)$ denote the estimation of $\alpha_0^*$, $\ve \alpha^*$, and $b^*(\cdot)$ at the $m$-th iteration, respectively.
We  approximate $\eta_i^*$ with $\eta_i^{(m)} = \alpha_0^{(m)} + {\bf z}_i^\intercal \ve \alpha^{(m)} + \int_{\mathcal{S}} x_i(s)b^{(m)}(s)ds$. Denote by
\begin{equation*}\label{score:fish}
r(y, \eta) = {\partial\over\partial \eta} \log p(y;\eta) = T(y) - \dot A(\eta), ~~~~w(\eta) = - \mathbb{E}\{{\partial^2\over \partial \eta^2} \log p(y;\eta) \mid x(\cdot), {\bf z}\} = \ddot A(\eta)
\end{equation*}
 the score function and Fisher information with respect to $\eta$, respectively,  where $\dot A(\eta)=dA(\eta)/d\eta$ and 
$\ddot A(\eta)=d^2A(\eta)/d\eta^2$. For ease of notation, let
$r_i^{(m)} = r(y_i, \eta_i^{(m)})$ and $w_i^{(m)} = w(\eta_i^{(m)})$. 
%To obtain $\ve \alpha^{(m+1)}$ and $b^{(m+1)}(\cdot)$, 
Motivated by the iteratively reweighted least squares (IRLS, \cite{green1984}),
we define the pseudo-response at the $m$-th iteration
\[
\tilde{y}_i^{(m)} = \eta_i^{(m)} + \{w_i^{(m)}\}^{-1} r_i^{(m)} ~\mbox{ for } i= 1, \ldots, n.
\]
Then, we can obtain $b^{(m+1)}(s)$ by applying the empirical RAPLS algorithm, which was presented above, to the PFLM
$\tilde{y}_i^{(m)} = \alpha_0 + {\bf z}_i^\intercal \ve \alpha + \int_{\mathcal{S}} x_i(s)b(s)ds + \epsilon_i^{(m)}$.
We then obtain $\alpha_0^{(m+1)}$ and
$\ve \alpha^{(m+1)}$ by solving
the score equation
\begin{equation}\label{alpha:mle}
%{\bf Z}^\intercal S_n^{(m+1)}(\ve \alpha) = {\bf 0},
\sum_{i=1}^n {\bf z}_i r(y_i, \alpha_0 + {\bf z}_i^\intercal \ve \alpha + \int_{\mathcal{S}}x_i(s) b^{(m+1)}(s)ds ) = {\bf 0}.
\end{equation}
We iterate this process until $|\alpha_0^{(m+1)} - \alpha_0^{(m)}| + \|\ve \alpha^{(m+1)} - \ve \alpha^{(m)}\| + \|b^{(m+1)}(s) - b^{(m)}(s)\| \leq \varrho$ for some pre-specificed small $\varrho$, say $10^{-4}$.

\begin{remark}
A similar  FPLS algorithm for the functional joint model (FJM), called  FJM-FPLS, has been used in our earlier work \cite{wang2020partial}. However, the population-level algorithm is missing in \cite{wang2020partial}, which hinders the theoretical justification of  FJM-FPLS. In the next section, we will take advantage of the population-level RAPLS algorithm in Section \ref{pop:RAPLS} to study the asymptotic properties of the empirical RAPLS algorithm.
\end{remark}

%\begin{remark}

%------------------------------------------------------------------------

\section{Theoretical Properties}\label{sec:conv:rate}

In this section, we establish the asymptotic properties of the estimators for GFLM in the previous section.  
%Similar results for the PFLM are provided in the supplementary document.

 Throughout the section, we consider the high-dimensional regime that the number of basis functions $p = p(n) \rightarrow \infty$ as $n \rightarrow \infty$. We make the following additional assumptions. 
 \begin{itemize}
     \item[(A1)] $\|b^*\| + \mathbb{E}(\|x\|^4) < \infty ~\mbox{and}~ \lambda_{\min}\left(\mathbb{E}({\bf z}^{\otimes 2})\right) > 0$, where $\lambda_{\min}({M})$ denotes the smallest eigenvalue of any symmetric matrix ${M}$. 
     \item[(A2)] $\|\mathcal{C}\| < 1$ and $p = O(n^{1/2})$ as $n \rightarrow \infty$.
  %  \item[(A4)] $\left\|\widehat{\theta}_k(s) - \theta_k^*(s)\right\| = o_p\left( n^{-1/4} \right)$.
 \end{itemize}

Assumption (A1) is a standard regularity condition. {The condition $\|\mathcal{C}\| < 1$ in Assumption (A2) holds by scaling the functional covariates $x(s)$ such that $\mathbb{E} \|x\|^4 < 1$. To see this, note that 
\[
\|K\|^2 = \iint_{\mathcal{S}^2} \left\{\mathbb{E} x(s) x(t) \right\}^2 dsdt \leq \mathbb{E} \|x\|^4 < 1. 
\]
Then, since $\mathcal{C}(s,t) = K(s,t) - \mathbb{E}\left\{{\bf z}^\intercal x(s)\right\} \left\{\mathbb{E} ({\bf z}^{\otimes 2})\right\}^{-1} \mathbb{E}\left\{{\bf z} x(t)\right\}$, we get $\|\mathcal{C}\| \leq \|K\| < 1$. } 
%We will discuss how to obtain $\widehat{\theta}_k(s)$ that satisfies (A4) later.   

%------------------------------------------------------------
%We next generalize Theorem \ref{thm:main1} to the GFLM (\ref{rev:1}).
%We first establish the convergence rate of any RAPLS iterate $b^{(m)}(\cdot)$.  
%and the asymptotic normality of $\widehat{\ve \alpha}_p^{\text{cal}}$ with the $\sqrt{n}$-convergence rate.
Recall that $\alpha_0^{(m)}$, $\ve \alpha^{(m)}$, and $b^{(m)}(\cdot)$, respectively, denote the $m$-th RAPLS iterate of $\alpha_0^*$, $\ve \alpha_p^*$, and $b^*(\cdot)$ under the GFLM (\ref{rev:1}) for $m \geq 0$. 
The following result shows that with deterministic initial values $\alpha_0^{(0)}, \ve \alpha^{(0)}, b^{(0)}(\cdot)$, the first-step iteration $b^{(1)}$ is not necessarily a consistent estimator of $b_p^*$.  
\begin{proposition}\label{prop1}
Suppose Assumptions (A1) and (A2) hold. Then, as $n \rightarrow \infty$, 
we have 
\(
\|b^{(1)} - b_p^*\| = O_p(\lambda_p^{-3}), 
\)
{
%\color{blue}
where $\lambda_p = \lambda_{\min}({\bf H}^*)$ is the smallest eigenvalue of ${\bf H^*}$, which is defined in \eqref{hbeta:true}. 
%Section \ref{pop:RAPLS}. 
}
\end{proposition}
%\begin{remark}
As $p \rightarrow \infty$ along with $n$, $\lambda_p$ may converge to 0, indicating that $b^{(1)}$ may not be a consistent estimator. 
%Specifically, the decay rate of $\lambda_p$ depends on the relationship between $\mathcal{C}(s,t)$ and $b^*(\cdot)$.
% how $b(\cdot)$ is aligned with the eigenvectors of $\mathcal{C}(s,t)$. To see th
% More specifically, recall from (\ref{hbeta:true}) that ${\bf H}^* = (h_{jk}^*)$, where 
% $
% h_{jk}^* = \int_{\mathcal{S}} \mathcal{C}^{j+1}(b^*)(s) \mathcal{C}^{k}(b^*)(s)ds = \iint_{\mathcal{S}^2} \mathcal{C}^{j}(b^*)(s) \mathcal{C}(s, t) \mathcal{C}^{k}(b^*)(t) dsdt.
% $ 
% Defining $\ve \Psi(s) = \left(\mathcal{C}(b^*), \ldots, \mathcal{C}^p(b^*)\right)^\intercal$, one can write 
% $ 
% {\bf H}^* =  \iint_{\mathcal{S}^2} \mathcal{C}(s,t) \ve \Psi(s) \ve \Psi(t)^\intercal dsdt. 
% $ 
% Let $\mathcal{C}(s,t) = \sum_{k} \mu_k \phi_k(s) \phi_k(t)$ denote the eigen-decomposition of $\mathcal{C}(s,t)$ and $b_k^* = \int_{\mathcal{S}} b^*(s)\phi_k(s)ds$. Then, some calculation yields that
% \(
% \mathcal{C}^j(b^*) = \sum_{k} \mu_k^j b_k^* \phi_k
% \)
% and
% \(
% {\bf H}^* = \sum_{k} \mu_k^3 {b_k^*}^2 {\ve \mu}_k {\ve \mu}_k^\intercal, 
% \)
% where $\ve \mu_k = (1, \mu_k, \ldots, \mu_{k}^{p-1})^\intercal$.
% This indicates that $\lambda_p$ depends on the eigenvalues of $\mathcal{C}(s,t)$ and $\{b_{k}^*\}_{k \geq 1}$,   characterizing the alignment between $b^*(\cdot)$ and the eigenvectors of $\mathcal{C}(s,t)$. 
% %\end{remark}
Thus, Proposition \ref{prop1} necessitates the use of better initial values to guarantee the consistency of the RAPLS estimates. 
%, which are consistent estimators of $\alpha_{0}^*, \ve \alpha^*$ and $b^*$. 
%obtained from some existing method such as FPCA. 
Letting $\alpha_{0,n}^{(0)}, \ve \alpha_n^{(0)}, b^{(0)}_n(\cdot)$ denote data-driven initial values with the convergence rate $\tau_n$, 
we introduce the following assumption. 
\begin{itemize}
\item[(A3)] As $n \rightarrow \infty$, $|\alpha^{(0)}_{0,n} - \alpha_0^*| + \|\ve \alpha^{(0)}_{n} - \ve \alpha^*\| + \|b^{(0)}_n - b^*\| = O_p(\tau_n)$ and $\|b_p^* - b^*\| = O(\lambda_p^{-2} \tau_n^2)$, 
{
%\color{blue} 
where $\tau_n = o(1)$ as $n \rightarrow \infty$. 
}
\end{itemize}
{
%\color{blue} 
Initial values that satisfy Assumption (A3) may be obtained from existing methods. For example, \cite{muller2005generalized} established $\sqrt{n}$-consistency for their functional estimator, constructed using arbitrary deterministic basis expansion. 
For PFLM, \cite{kong2016partially} used FPCA and penalized approaches, and established consistency of their estimators where the convergence rate depends on the eigenstructure of the covariance kernel. 
Additionally, \cite{lv2023kernel} constructed an estimator within a reproducing kernel Hilbert space (RKHS) and established its convergence rate as a function of the decay rate of the eigenvalues of the ``orthonormalized" covariance kernel. 
The theorems below require $\lambda_p^{-2}\tau_n = O(1)$ or $\lambda_p^{-2}\tau_n^2 = o(n^{-1/4})$. 
These two assumptions are generally weak
as $\lambda_p$ usually converges to 0 at a slow rate. 
Thus, all the above methods can be used to obtain initial values that satisfy the required assumptions.
}

{
%\color{blue}
While Assumption (A3) seems a strong assumption,  we show in the following result that our RAPLS iterates
can potentially achieve faster convergence rates than the initial values. 
Furthermore, in finite-sample applications, we will illustrate that
%In practice, we recommend running the RAPLS algorithm until the iterations become stable to achieve more robust performance. 
%While Theorem \ref{thm:main2} requires consistency of the initial values  $\alpha_0^{(0)}, \ve \alpha^{(0)},$ and $b^{(0)}(\cdot)$, 
in Sections \ref{sec:simu} and \ref{sec:real}, the RAPLS algorithm with simple deterministic initial values, such as $\alpha_0^{(0)} = 0, \ve \alpha^{(0)} = {\bf 0}$, and $b^{(0)}(\cdot) \equiv 0$, or random initial values, 
can still outperform existing methods in terms of estimation and prediction accuracy. 
}

%------------------------------------------------------------
%%%%% split Theorem 1 into two parts 

\begin{theorem}\label{thm:main2}
Suppose Assumptions (A1)--(A3) hold. If $\lambda_p^{-2} \tau_n = O(1)$ as $n \rightarrow \infty$, then
for each $m \geq 1$, we have
\begin{align}\label{thm2:eq1}
%\label{thme:est:b}
  \left\|b^{(m)} - b^*\right\| =  O_p\left( 
 \lambda_p^{-2} \tau_n^2 
 \right).% \mbox{ as } n \rightarrow \infty.
\end{align}
%----------------------------------------------
%%% split from here ----
% Furthermore, if Assumptions (A1)-(A5) hold and $\tau_n^2\lambda_p^{-2} = o(n^{-1/4})$, we have, as $n\rightarrow \infty$,
% \begin{equation}\label{thm2:eq2}
% \sqrt{n}(\widehat{\ve \alpha}^{\text{cal}}_{p} - \ve \alpha^*) \stackrel{d}{\rightarrow} N\left(0, \widetilde{\Sigma}_\zeta^{-1}\right), % \mbox{ as } n \rightarrow \infty,
% \end{equation}
% where $\widetilde{\Sigma}_\zeta$ is defined in Assumption (A1).
\end{theorem}
Eq. (\ref{thm2:eq1}) implies that 
%the initial value is within a bounded neighborhood of the true parameters and 
if $\lambda_p^{-2} \tau_n = O(1)$, then all RAPLS iterates are consistent estimators with the convergence rate $O_p(\tau_n)$. 
%Furthermore, if $\lambda_p^{-2}\tau_n = o(1)$ as $n \rightarrow \infty$,
In particular, if $\lambda_p^{-2}\tau_n = o(1)$, 
$b^{(m)}$ converges to $b^*$ at a faster rate than the initial value $b_n^{(0)}$ for all $m \geq 1$. 
%which is likely when the number of basis functions $p$ is much smaller than the sample size $n$. 
%Note that consistency of $b^{(m)}$ does not require the linear relation between ${\bf z}_i$ and $x_i(\cdot)$. 
%Note that when $\lambda_p$ is exactly zero, Theorem \ref{thm:main2} becomes less meaningful. 
{
%\color{blue}
When $\lambda_p$ is a constant, 
Theorem \ref{thm:main2} simplies to $\|b^{(m)} - b^*\| = O_p(\tau_n^2)$. Since $\tau_n = o(1)$, this indicates $b^{(m)}$ has a much faster convergence rate than the initial values. 
%can further simplify when $\lambda_p$ is constant. 
This special situation occurs if there exists a constant $M_\lambda$ such that either (i) $\theta_k = 0$ for $k > M_\lambda$, or (ii) $b_k^* = 0$ for $k > M_\lambda$. Scenario (i) arises if $x_i(\cdot)$ belongs to the $M_\lambda$-dimensional space $\mathcal{S}(M_\lambda)$ for each $i = 1, \ldots, n$, while scenario (ii) holds if $b^*(\cdot)$ belongs to $\mathcal{S}(M_\lambda)$.
%the space spanned by $\phi_1(\cdot), \ldots, \phi_{M_\lambda}(\cdot)$. 
%In both scenarios, one can see that the rank of ${\bf H}^*$ is at most $M_\lambda$; thus, $\lambda_p$ would be $0$ for all $p > M_\lambda$. 
% In both scenarios, the infinite-dimensional GFLM simplifies to a finite-dimensional parametric model, and Theorem \ref{thm:main2} would require weaker conditions. 
 %Therefore, Theorem \ref{thm:main2} still holds for $p \leq \mbox{rank}({\bf H}^*)$ for these two scenarios. 
}

%----------------------------------------------------------------

It is unfortunate that the ``plug-in" estimators in (\ref{alphaest:pflm}) and (\ref{alpha:mle}) are not asymptotically normal nor efficient. 
%due to their ``plug-in" nature. 
This issue is common in semiparametric inference of the low-dimensional parameter in the presence of high-dimensional nuisance parameters, where the plug-in approach causes a potential bias and fails to be efficient \citep{victor17}. 
%--------------------------------------------------------------
{
%\color{blue}
To address this issue, we develop a calibration procedure. 
First, recall that $w(\eta)$ is the information function with respect to the linear predictor $\eta$ for the GFLM \eqref{rev:1}. 
We introduce a new functional operator $K_w: f \rightarrow \int K_w(s,t) f(s)ds$, where
$K_w(s, t) = \mathbb{E}[w^* x(s) x(t)]$ with $w^* = w(\eta^*)$. 
Since $w^* > 0$, $K_w(s, t)$ represents a generalized covariance kernel, and thus is always positive definite. 
%; more discussions of $K_w(s, t)$ can be found in \cite{muller2005generalized}. 
Then, we define $\zeta_{ik} = z_{ik} - \int_{\mathcal{S}} x_i(s) K_w^{-1} ( \mathbb{E}[w_i^* x_i(s) z_{ik} ] ) ds$ for $i = 1, \ldots, n$ and $k = 1, \ldots, q$, where
$K_w^{-1}$ is the inverse operator of $K_w$. 

One can check that 
\begin{align*}
\mathbb{E}[w_i^* x_i(s)\zeta_{ik}] = &~ \mathbb{E}[w_i^* x_i(s) z_{ik}] - \mathbb{E}[w_i^* x_i(s) \int_{\mathcal{S}} x_i(t) K_w^{-1} ( \mathbb{E}[w_i^* x_i(t) z_{ik} ] ) dt ] \\
= &~ \mathbb{E}[w_i^* x_i(s) z_{ik}] - \int_{\mathcal{S}} \mathbb{E}[w_i^*x_i(s)x_i(t)] K_w^{-1} ( \mathbb{E}[w_i^* x_i(t) z_{ik} ] ) dt \\
= &~ \mathbb{E}[w_i^* x_i(s) z_{ik}] - K_w \left( K_w^{-1} ( \mathbb{E}[w_i^* x_i(s) z_{ik} ] ) \right) \\
= &~ 0.
\end{align*}
%$\mathbb{E}[w_i^* \zeta_{il} x_i(s)] = 0$ 
for each $i, l$ and $s \in \mathcal{S}$.
For ease of notation, denote $\theta^*_{k}(s) = K_w^{-1} ( \mathbb{E}[w_i^* x_i(s) z_{ik} ] )$ for $k=1, \ldots, q$, and 
Let $\widehat{\theta}_{k}(s)$ being an estimate of $\theta^*_{k}(s)$ that will be discussed below. 
We further calculate 
$\widehat{\zeta}_{ik} = z_{ik} - \int_{\mathcal{S}} x_i(s) \widehat{\theta}_k(s) ds$ for $i = 1, \ldots, n$ and $k = 1, \ldots, q$.
%$\widehat{{\ve \theta}}(s) = (\widehat{\theta}_{1}(s), \ldots, \widehat{\theta}_{q}(s))^\intercal$ 
% with $\widehat{\theta}_{k}(s)$ being an estimate of $K_w^{-1} ( \mathbb{E}[w_i^* x_i(s) z_{ik} ] )$ for $k = 1, \ldots, q$, 
%With some ``good" estimate $\widehat{\zeta}_{ik}$ that will be discussed in detail in the next section, 
Finally, the calibrated estimator of $\ve \alpha$ is defined as
\begin{align}\label{alpha:cal:gflm}
    \widehat{\ve \alpha}_p^{\text{cal}} = \argmax_{\ve \alpha} n^{-1}\sum_{i=1}^n \log p\left(y_i, \int_{\mathcal{S}} x_i(s) \left\{ \widehat{b}_p(s) + \widehat{\ve \alpha}_p^\intercal \widehat{\ve \theta}(s) \right\} ds + \widehat{\alpha}_{0,p} + \widehat{\ve \zeta}_i^\intercal \ve \alpha \right),
\end{align}
where $\widehat{\ve \zeta}_i = (\widehat{\zeta}_{i1}, \ldots, \widehat{\zeta}_{iq})^\intercal$, 
$\widehat{\ve \theta}(s) = (\widehat{\theta}_1(s), \ldots, \widehat{\theta}_q(s))^\intercal$, and 
$\widehat{b}_p(s)$, $\widehat{\alpha}_{0,p}$, and $\widehat{\ve \alpha}_p$ are, respectively, the $p$-th order RAPLS estimates of $b^*(\cdot)$,  $\alpha_0^*$ and $\ve \alpha^*$ at convergence. 
%A similar calibration procedure for the PFLM  is provided in the supplementary document. 
}

%----------------------------------------------------------------
{
%\color{blue}
We next discuss how to obtain each $\widehat{\theta}_k(\cdot)$. 
%the theoretical property of the calibrated scalar estimator $\widehat{\ve \alpha}_p^{\text{cal}}$. 
%Recall from Section \ref{sec:3} that $\zeta_{ik} = z_{ik} - \int_{\mathcal{S}} x_i(s) K_w^{-1} ( \mathbb{E}[w_i^* x_i(s) z_{ik} ] ) ds$ for $i=1, \ldots, n$ and $k=1, \ldots, q$. 
%For ease of notation, denote $\theta^*_{k}(s) = K_w^{-1} ( \mathbb{E}[w_i^* x_i(s) z_{ik} ] )$ for $k=1, \ldots, q$. 
%To estimate each $\zeta_{il}$, we only need to estimate each $\theta^*_{l}(s)$. 
%We conclude this section with discussions on the estimation of $\theta_k^*(\cdot)$ for nonlinear GFLMs. Note that under Assumption (A1), existing methods for SFLMs are no longer appropriate for estimating $\theta_k^*(\cdot)$.  
Motivated by the fact that 
\(
\mathbb{E}[w_i^* \zeta_{ik} x_i(s)] = 0, 
\)
we develop a three-step estimation procedure for each $\zeta_{ik}$. 

{\bf Step 1:} We estimate $w_i^*$ with $\widehat{w}_i = w(\widehat{\eta}_i)$ respectively, where $\widehat{\eta}_i = \widehat{\alpha}_{0,p} + {\bf z}_i^\intercal\widehat{\ve \alpha}_{p} + \int_{\mathcal{S}} x_i(s) \widehat{b}_p(s)ds$ with $\widehat{\alpha}_{0,p}, \widehat{\ve \alpha}_{p},$ and $\widehat{b}_p(s)$ obtained from the RAPLS algorithm. 

{\bf Step 2:} With a set of deterministic orthonormal basis functions $\{\pi_j(\cdot)\}_{j\geq 1}$,  we expand $x_i(s)$ and $\theta(s)$, respectively, as 
\[
x_i(s) = \sum_{j=1}^ \infty U_{ij} \pi_j(s) \mbox{ and } \theta_k^*(s) = \sum_{j=1}^\infty \theta_{kj}^* \pi_{j}(s), 
\]
where $U_{ij} = \int x_i(s)\pi_j(s)ds$ and $\theta_{kj}^* = \int \theta^*_k(s)\pi_j(s)ds$ for $k = 1, \ldots, q$ and
all $i$ and $j$. With a pre-selected truncation parameter $s_n$, 
%one can estimate $\theta^*_{k1}, \ldots, \theta^*_{k,s_n}$ by minimizing
we obtain the weighted least squares estimator of $\{\theta_{kj}^*\}$: 
\[
\{\widehat{\theta}_{k1}, \ldots, \widehat{\theta}_{ks_n}  \} = \mbox{argmin}_{\theta_{k1}, \ldots, \theta_{ks_n}} \sum_{i=1}^n \widehat{w}_i (z_{ik} - \sum_{j=1}^{s_n} U_{ij} \theta_{kj})^2 \mbox{ for } k = 1, \ldots, q.
\]
% Some algebra yields that 
% \(
% \widehat{\ve \theta}_{k} = (\widehat{\theta}_{k1}, \ldots, \widehat{\theta}_{k,s_n})^\intercal = 
% (U_{s_n}^\intercal \widehat{A} U_{s_n})^{-1} U_{s_n}^\intercal \widehat{A} Z_k, 
% \)
% where $\widehat{A} = \mbox{diag}\left(\widehat{w}_1, \ldots, \widehat{w}_n \right)$, $U_{s_n} = \left( U_{ij} \right)_{i = 1, \ldots, n; j = 1, \ldots, s_n}$ and $Z_k = (z_{1k}, \ldots, z_{nk})^\intercal$. 

{\bf Step 3:}
We obtain $\widehat{\theta}_{k}(s) = \sum_{j=1}^{s_n} \widehat{\theta}_{kj} \pi_j(s)$ and $\widehat{\ve \zeta}_i = (\widehat{\zeta}_{i1}, \ldots, \widehat{\zeta}_{iq})^\intercal$, where
$\widehat{\zeta}_{ik} = z_{ik} - \int_{\mathcal{S}} x_i(s) \widehat{\theta}_{k}(s) ds$ for $i =1, \ldots, n$ and $k = 1, \ldots, q$. 
}
 
Plugging $\widehat{\ve \zeta}_{i}$ into \eqref{alpha:cal:gflm}, we obtain the calibrated estimator $\widehat{\ve \alpha}_p^{\text{cal}}$.  
%according to \eqref{alpha:cal:gflm}. 
The following result establishes the asymptotic normality of $\widehat{\ve \alpha}_p^{\text{cal}}$. 
\begin{theorem}\label{thm:main3}
Suppose Assumptions (A1)--(A3) hold. If
\( \tau_n^2\lambda_p^{-2} = o(n^{-1/4}), 
\|\theta_k^*(s) - \sum_{j=1}^{s_n} \theta_{kj}^* \pi_j(s)\|^2 = O(s_n^{1-2b}), 
\) 
where $s_n = C n^a$
for some constant $C$ 
with $1/\{2(2b-1)\} < a \leq 1/4$, 
then we have
%\begin{align}\label{thm2:eq1}
% %\label{thme:est:b}
%   \left\|b^{(m)} - b^*\right\| =  O_p\left( 
%  \lambda_p^{-2} \tau_n^2 
%  \right).% \mbox{ as } n \rightarrow \infty.
% \end{align}
%----------------------------------------------
%%% split from here ----
% Furthermore, if Assumptions (A1)-(A5) hold and $\tau_n^2\lambda_p^{-2} = o(n^{-1/4})$, we have, 
as $n\rightarrow \infty$,
\begin{equation}\label{thm2:eq2}
\sqrt{n}(\widehat{\ve \alpha}^{\text{cal}}_{p} - \ve \alpha^*) \stackrel{d}{\rightarrow} N\left(0, {\Sigma}_\zeta^{-1}\right), % \mbox{ as } n \rightarrow \infty,
\end{equation}
where ${\Sigma}_\zeta = \mathbb{E} \left\{ w_i^* {\ve  \zeta}_i{\ve \zeta}_i^\intercal \right\}$. 
\end{theorem}
{
%\color{blue}
Theorem \ref{thm:main3} imposes a slightly different condition on $\tau_n$ compared to Theorem \ref{thm:main2}, though neither condition is necessarily stronger than the other. 
Specifically, 
when $\tau_n = o(n^{-1/4})$, the condition $\tau_n \lambda_p^{-2} = O(1)$ from Theorem \ref{thm:main2} implies the condition in Theorem \ref{thm:main3}. 
Conversely, when $\tau_n \gtrsim n^{-1/4}$, the condition $\tau_n^2 \lambda_p^{-2} = o(n^{-1/4})$ from Theorem \ref{thm:main3} implies the condition in Theorem \ref{thm:main2}. 
% under the condition $\tau_n \lambda_p^{-2} = O(1)$ from Theorem \ref{thm:main2}, we obtain 
% $\tau_n^2\lambda_p^{-2} = o(n^{-1/4})$ if $\tau_n = o(n^{-1/4})$. 
% Conversely, 
% under the condition $\tau_n^2\lambda_p^{-2} = o(n^{-1/4})$ from Theorem \ref{thm:main3}, if $\tau_n \gtrsim n^{-1/4}$, we can also satisfy $\tau_n \lambda_p^{-2} = O(1)$. 

The condition $\|\theta^*_{k}(s) - \sum_{k=1}^{s_n} \theta_{kj}^* \pi_j(s)\|^2 = O(s_n^{1-2b})$ requires the $s_n$-term approximation of $\theta^*_{k}(s)$ based on the basis functions $\{\pi_j(\cdot)\}$ to be sufficiently accurate. 
Since $s_n = Cn^{a}$ and $1/\{2(1-2b)\} < a$, we can derive that 
\[
\|\theta^*_{k}(s) - \sum_{k=1}^{s_n} \theta_{kj}^* \pi_j(s)\| = O(n^{a(1-2b)/2}) = o(n^{-1/4}),
\]
which is a relatively slow convergence rate. 

Theorem \ref{thm:main3} also implies the semi-parametric efficiency of the calibrated estimator $\widehat{\ve \alpha}_p^{\text{cal}}$. 
In practice, ${\Sigma}_\zeta$ can be estimated by 
$\widehat{\Sigma}_\zeta = n^{-1} \sum_{i=1}^n \widehat{w}_i \widehat{\ve \zeta}_i \widehat{\ve \zeta}_i^\intercal$, where
$\widehat{w}_i$ and $\widehat{\ve \zeta}_i$ are, respectively, estimated in Step 1 and Step 3. 
% With $\widehat{\Sigma}_\zeta$, one can perform an asymptotic test for testing $H_0: {\bf B}\ve \alpha^* = \ve 0$ vs. $H_1: {\bf B}\ve \alpha^* \neq \ve 0$, where ${\bf B} \in \mathbb{R}^{l \times q}$ is an arbitrary contrast matrix. 
% Specifically, the test statistic is 
% \[
% T = (\widehat{\ve \alpha}^{\text{cal}}_{p} - \ve \alpha^*)^\intercal {\bf B}^\intercal (\sum_{i=1}^n \widehat{w}_i \widehat{\ve \zeta}_i \widehat{\ve \zeta}_i^\intercal)^{-1} {\bf B} (\widehat{\ve \alpha}^{\text{cal}}_{p} - \ve \alpha^*) \stackrel{d}{\rightarrow} \chi^2_{l}. 
% \]
}

% We prove in Lemma 7 of the supplementary document that $\widehat{\theta}_k(s)$ satisfies Assumption (A4) if
% \( \tau_n^2\lambda_p^{-2} = o(n^{-1/4}), 
% \theta_{kj}^* \leq Cj^{-b}, s_n \approx n^a, b > 3/2, \mbox{ and } 1/\{2(2b-1)\} < a \leq 1/4.
% \)
% One may replace the deterministic basis functions $\{\pi_k(\cdot)\}_{k \geq 1}$ with data-driven basis functions such as FPCA or FPLS basis functions. However, this may require different assumptions and proofs, which are not discussed here.  

%--------------------------------------------------------------

%Since $x_i(\cdot)$ is ``orthogonal" to ${\ve \zeta}_i$ in  (\ref{add:irls:gflm:2})  in the sense that $\mathbb{E}\left[ w_i^* x_i(s) {\ve \zeta}_i \right] = {\bf 0}$ for any $s$, $\widetilde{\Sigma}_\zeta^{-1}$ is a lower bound on the asymptotic variance of a regular estimator of $\ve \alpha^*$ in model (\ref{rev:1}), which is achieved by $\widehat{\ve \alpha}^{\text{cal}}_{p}$, as shown in (\ref{thm2:eq2}). More precisely, one can establish the asymptotic efficiency of $\widehat{\ve \alpha}^{\text{cal}}_{p}$ among all regular estimators of $\ve \alpha^*$ by following the definition of ``regular estimators" in \cite{begun1983information}.

%--------------------------------------------------------
\section{Simulation Studies}\label{sec:simu}

{
%\color{blue}
We examined the finite-sample performance of the proposed RAPLS method in two settings: a partially functional linear model (PFLM) and a functional Poisson model (FPM). 
We compared RAPLS with two existing methods: the functional principal component regression (FPCR, \citealp{hall2007methodology}), and an existing method of partial least squares for generalized linear models (plsRglm, \citealp{meyer2010comparaison}). %and the reproducing kernel Hilbert space for generalized linear models (RKHSglm, \citealp{yuan2010reproducing, kadri2010nonlinear}).
As a key procedure in FPCR, FPCA was performed using the R package {\tt fdapace}; plsRglm was performed using the R package {\tt plsRglm}. 
%As no public code is available for RKHSglm, we wrote our own code to implement RKHSglm based on the theory in \cite{yuan2010reproducing, kadri2010nonlinear} and the R package {\tt glmnet}. 
% In the PFLM setting, the generalized cross validation (GCV) was used to select 
% the optimal number of RAPLS basis functions and FPCs, while 
% in the FPM setting, AIC was used. 
AIC was used to determine the optimal number of components for all the methods. 
% and the optimal tuning parameter for RKHSglm was determined by 5-fold cross validation, implemented by {\tt cv.glmnet()}. 
% Following \cite{yuan2010}, the RKHS was selected as $\mathcal{H} = \mathcal{W}_2^2$ with the associated penalty $J(\beta) = \int (\ddot{\beta})^2$; we refer the readers to Section 5 of \cite{yuan2010} for more details. 
}

{
%\color{blue} 
In both settings, each curve $x_i(t)$ was generated in a way similar to that in \cite{yuan2010}. More specifically, for $t \in [0,1]$ and $k = 1, \ldots, 50$, let $
\phi_{k}(t) = \sqrt{2}\cos(k\pi t)$.  
%and $\phi_{2k}(t) = \sqrt{2}\sin(k\pi t)$. 
 For $i = 1, \ldots, n$, we generated $x_i(t)$ according to $x_i(t) = \sum_{k=1}^{50} k^{-1/4} \xi_{ik} \phi_k(t)$, where $\xi_{ik} \stackrel{i.i.d}{\sim} N(0,1)$. 
 These curves were evaluated at 900 equally spaced points on $[0,1]$. 
 Simple algebra yields that 
 \(
 \mbox{Cov}(x_i(s), x_i(t)) = \sum_{k=1}^{50} k^{-1/2} \phi_k(s) \phi_k(t), 
 \)
 indicating that the eigenvalues are $1, 1/\sqrt{2}, \ldots, 1/\sqrt{50}$, and the eigenfunctions are $\phi_1(\cdot), \ldots, \phi_{50}(\cdot)$. 
 %, and generated $z_i$ from $N(0, \xi_{i5}^2)$. 
%We took observations for each $x_i(s)$ at $s = 1/900, 2/900, \ldots, 1$. Similar to Setting I, we considered two scenarios of $b(s)$: (A) $b(s) = \phi_1(s) + \phi_2(s) + \phi_3(s)$ and (B) $b(s) = \phi_4(s) + \phi_5(s) + \phi_6(s)$. Finally, for $i = 1, \ldots, n$, we generated the outcome $y_i = 0.5 + z_i + \int_0^1 x_i(s)b(s)ds + \delta_i$, where $\delta_i \sim N(0,1)$. 
Finally, the scalar covariate $z_i$ was generated from a normal distribution with mean $0$ and variance $\xi_{i5}^2/5$, imposing correlations between $z_i$ and $x_i(\cdot)$. 
}

%----------------------------------------------------------------
%%%%%%%%%%%%%%%%%%%%%%%%%%%%%%%%%%%%%%%%%%%%%%%%%%%%%%%%%%%%%%%%%

\subsection{Setting I: PFLM}\label{sec:4.2}

In this setting, given $\{x_i(\cdot), z_i\}_{i=1, \ldots, n}$, the outcome $y_i$ was generated according to 
\(
y_i = 0.5 + \alpha^* z_i + \int_{0}^1 x_i(s)b^*(s)ds + \epsilon_i,
\)
where $\epsilon_i$ follows a normal distribution with mean 0 and variance $0.8$.
We considered $\alpha^* = 1$ and two scenarios of $b^*(\cdot)$. 
In the first scenario, $b^*(\cdot)$ was in the span of the top $25$ eigenfunctions as 
$b^*(s) = \sum_{k=1}^{25} (-1)^k \phi_k(s)$ for $s \in [0,1]$. 
In the second scenario, $b^*(\cdot)$ was in the span of the tail $25$ eigenfunctions as 
$b^*(s) = \sum_{k=26}^{50} (-1)^k \phi_k(s)$ for $s \in [0,1]$. 
Theoretically, $b^*(\cdot)$ in the first scenario is easier to estimate, as top eigenfunctions are easier to estimate than tail eigenfunctions.

We generated 500 independent data sets for $n = 100, 200$ and $500$. 
% For each data set, generated images $\{x_i(s)\}_{i=1}^n$ were unfolded to obtain vectors of size $d = 300 \times 300 = 90000$. We then fitted the functional Poisson model using both  RAPLS and FPCA with the optimal number of RAPLS/FPCA basis functions determined by the Akaike information criterion (AIC, \cite{akaike1998information}).
% The RAPLS algorithm was implemented using the initial value 0, and FPCA was implemented using the method based on the singular value decomposition proposed by \cite{zipunnikov2011functional}. 
%FPCA was implemented in the same way as Setting I. 
For each replication, we implemented RAPLS, FPCR, and plsRglm, and obtained the estimates of $b^*(\cdot)$.
We implemented the proposed calibration procedure
to estimate the scalar parameter $\alpha^*$. 
%, referred to as RAPLS-cal hereafter, and compared it the FPCR and plsRglm estimates. 
% The calibrated estimator of $\alpha$, referred to as RAPLS-cal hereafter,  was obtained by fitting $z_i = \int_{\mathcal{S}} x_i(s) \theta(s)ds + \zeta_i$ using the proposed two-step procedure with Fourier basis functions. 
To account for both bias and variance, we reported mean squared errors for all these estimates. 
%We first examined the estimation and prediction accuracy using the MSE and PMSE according to 
Specifically, we calculated 
$\int_0^1 (\widehat{b}(s)-b^*(s))^2ds$ 
and 
$(\hat{\alpha} - \alpha^*)^2$ for all estimators, and reported their averages over 500 replications as $\mbox{MSE} (b)$ and  $\mbox{MSE} (\alpha)$, respectively. 
% and $\mbox{PMSE} = \mathbb{E}_{x^*} \left[ \int_{\mathcal{S}} x^*(s)\{\widehat{b}(s)-b(s)\} ds \right]^2$,  where $x^*(\cdot)$ is an independent copy of $x(\cdot)$.
%-----------------------
%%%%% TBA -----
%%%%%% Results TBA ----
%  \begin{figure}[t!]
%   \centering
% \includegraphics[width = 0.9\textwidth,height = 0.5\textheight]{R1_simu_linear_BIC.pdf}
% \caption{Simulation results: boxplots of $\mbox{MSE}(b)$ over 500 independent data sets for $n = 100, 200, 500$: Scenario I corresponds to $b^*(s) = \sum_{k=1}^{25} (-1)^k \phi_k(s)$, while Scenario II corresponds to $b^*(s) = \sum_{k=26}^{50} (-1)^k \phi_k(s)$. }\label{linear}
% \end{figure}
% {\color{blue}

{
%\color{blue}
The results for $\mbox{MSE}(b)$ are presented in Table \ref{linear}. In both settings, the proposed RAPLS method consistently outperforms FPCR and plsRglm in estimating the functional parameter $b^*$, highlighting the effectiveness of RAPLS in the PFLM context. While plsRglm is also based on partial least squares, it performs the worst in nearly all settings, except for Scenario II with $n = 100$. This is because, unlike RAPLS, which efficiently handles the correlation between scalar and functional covariates when constructing PLS basis functions, plsRglm simply combines the scalar and functional covariates to calculate the PLS components. As a result, the extracted plsRglm components fail to effectively represent the relationship between the functional covariates and the outcome. FPCR, on the other hand, performs worst in Scenario II with $n = 100$, underscoring a major limitation of FPCR/FPCA: when the true functional parameter is aligned with tail eigenvectors, FPCA struggles to estimate these eigenvectors in small sample sizes, leading to poor functional parameter estimation.
Notably, RAPLS demonstrates similar performance across both scenarios of $b^*(\cdot)$, highlighting its robustness regardless of the alignment between $b^*(\cdot)$ and the eigenfunctions.

}

\begin{table}[h]
\centering
\begin{tabular}{ccccc}
\hline
$n$ & Scenario & RAPLS & FPCR & plsRglm\\
\hline
100 & I & 0.834 (0.28) & 2.109 (1.06) & 4.440 (1.18)\\
200 & I & 0.263 (0.06) & 0.701 (0.52) & 2.487 (0.65)\\
500 & I & 0.089 (0.02) & 0.344 (0.31) & 1.545 (0.35)\\
100 & II & 0.820 (0.27) & 2.853 (1.68) & 1.884 (0.48)\\
200 & II & 0.265 (0.06) & 1.137 (1.03) & 1.272 (0.35)\\
%\addlinespace
500 & II & 0.089 (0.02) & 0.725 (0.87) & 1.169 (0.25)\\
\hline
\end{tabular}
\caption{Simulation results for the PFLM: $\mbox{MSE}(b)$ and the standard error over 500 independent data sets for $n = 100, 200, 500$: Scenario I corresponds to $b^*(s) = \sum_{k=1}^{25} (-1)^k \phi_k(s)$, while Scenario II corresponds to $b^*(s) = \sum_{k=26}^{50} (-1)^k \phi_k(s)$.}
\label{linear}
\end{table}

%----------------------------------------------------------
{
%\color{blue}
 We also compared RAPLS, FPCR, and plsRglm estimators of $\alpha^*$ with results presented in Table \ref{linear:alpha}. 
 The proposed calibrated estimator achieves the best estimation accuracy in all scenarios. 
 %both scenarios when $n = 100$ and 200. When $n = 500$, plsRglm performs slightly better than RAPLS-cal, despite plsRglm's poor performance in estimating $b^*$. 
 %Notably, even without calibration, RAPLS outperforms FPCR in terms of estimating $\alpha^*$, and the improvement from RAPLS to RAPLS-cal appears to be incremental. 
 }
 % Since RAPLS had convergence issues in a few replications, as a more robust alternative, we considered the median absolute deviation (MAD) according to $\mbox{MAD}(\widehat{\alpha}) = \widehat{\mbox{median}} \left( |\widehat{\alpha} - \alpha| \right)$ for any given estimator $\widehat{\alpha}$.
 \begin{table}[h]
 \centering
 \begin{tabular}{ccccc}
\hline
$n$ & Scenario & RAPLS & FPCR & plsRglm\\
\hline
100 & I & 0.0099 (0.016) & 0.0143 (0.021) & 0.0109 (0.018)\\
200 & I & 0.0027 (0.003) & 0.0034 (0.004) & 0.0028 (0.003)\\
500 & I & 0.0008 (0.001) & 0.0009 (0.001) & 0.0009 (0.001)\\
100 & II & 0.0099 (0.016) & 0.0152 (0.023) & 0.0114 (0.019)\\
200 & II &  0.0027 (0.003) & 0.0043 (0.007) & 0.0029 (0.003)\\
%\addlinespace
500 & II & 0.0008 (0.001) & 0.0010 (0.001) & 0.0009 (0.001)\\
\hline
\end{tabular}
\caption{Simulation results for the PFLM: $\mbox{MSE}(\alpha)$ over 500 independent data sets for $n = 100, 200, 500$: Scenario I corresponds to $b^*(s) = \sum_{k=1}^{25} (-1)^k \phi_k(s)$, while Scenario II corresponds to $b^*(s) = \sum_{k=26}^{50} (-1)^k \phi_k(s)$.}
\label{linear:alpha}
\end{table}

% The results in Table \ref{poisson:alpha} show that the RAPLS estimator outperforms FPCA. The additional one-step calibration further improves estimation accuracy. Notably, RAPLS-cal leads to similar estimation accuracy for both scenarios, whereas the other three methods yield substantially higher accuracy for Scenario A where $b(\cdot)$ aligns with top eigenfunctions. This indicates the effectiveness of the proposed calibration procedure regardless of $b(\cdot)$.

%------------------------------------------------------------------
%   \begin{figure}[t!]
%   \centering
% \includegraphics[width = 0.9\textwidth,height = 0.4\textheight]{poisson_est_pred.eps}
% \caption{Simulation results: boxplots of the MSE and PMSE results for Scenarios A and B over 500 independent data sets. Both RAPLS and FPCA methods are implemented based on the optimal number of basis functions determined by AIC.}\label{poisson}
% \end{figure}

%%%%%%%%%%%%%%%%%%%%%%%%%%%%%%%%%%%%%%%%%%%%%%%%%%%%%%%%%%%%%%%%%%%
\subsection{Setting II: FPM}
% We compared the proposed RAPLS with the NIAPLS method (an existing FPLS approach, \cite{kramer2011degrees}) and the reproducing kernel Hilbert space (RKHS) approach \citep{yuan2010} in a new PFLM setting. 
% The NIAPLS method was selected among various existing FPLS approaches since it has been implemented in the existing R package {\tt fda.usc}.
% In this setting, we considered a lower-dimensional case such that each curve $x_i(s)$ was represented by a $900 \times 1$ vector in a setting similar to the simulation study in \cite{yuan2010}. More specifically, for $t \in [0,1]$ and $k = 1,2,3$, let $\phi_{2k-1}(t) = \sqrt{2}\cos(2k\pi t)$ and $\phi_{2k}(t) = \sqrt{2}\sin(2k\pi t)$. For $i = 1, \ldots, n$, we generated $x_i(s)$ according to $x_i(s) = \sum_{k=1}^6 k^{-1/4} \xi_{ik} \phi_k(s)$, in which $\xi_{ik} \stackrel{i.i.d}{\sim} N(0,1)$, and generated $z_i$ from $N(0, \xi_{i5}^2)$. We took observations for each $x_i(s)$ at $s = 1/900, 2/900, \ldots, 1$. Similar to Setting I, we considered two scenarios of $b(s)$: (A) $b(s) = \phi_1(s) + \phi_2(s) + \phi_3(s)$ and (B) $b(s) = \phi_4(s) + \phi_5(s) + \phi_6(s)$. Finally, for $i = 1, \ldots, n$, we generated the outcome $y_i = 0.5 + z_i + \int_0^1 x_i(s)b(s)ds + \delta_i$, where $\delta_i \sim N(0,1)$. 
For $i = 1, \ldots, n$, the outcome $y_i$ was generated from a Poisson distribution with parameter $\exp(\eta_i)$, where 
$\eta_i = 0.5 + \alpha^* z_i + \int_0^1 x_i(s)b^*(s)ds$. 
Like Setting I, we considered $\alpha^* = 1$ and two scenarios of $b^*(\cdot)$. 
However, the choices of $b^*(\cdot)$ in Setting I were too large in scale, which led to infinite values in $y_i$. 
Thus, we rescaled each $b^*(\cdot)$ as $b^*(s) = \frac{2}{3} \sum_{k=1}^{25} (-1)^k \phi_k(s)$ for the first scenario and $b^*(s) = \frac{2}{3} \sum_{k=26}^{50} (-1)^k \phi_k(s)$ for the second scenario. 
{
%\color{blue}
To illustrate the robustness of the proposed RAPLS method regarding the initial values, 
we considered both deterministic and random initial values for RAPLS. 
For deterministic initial values, we set $b^{(0)}(\cdot) \equiv 0$, and then obtained $\alpha^{(0)}$ fitting the Poisson regression function with $y_i$ as the outcome and $z_i$ as the covariate. 
}
{
%\color{blue}
%We also considered different 
For random initial values, 
%and the results are reported in Table xxx. 
 $b^{(0)}(\cdot)$ was generated from a Gaussian process with zero mean and the kernel $K(s,t) = e^{-10(s-t)^2}$ for each replication. 
Then, $\alpha^{(0)}$ was obtained by fitting the Poisson regression model where $y_i$ is the outcome, $z_i$ is the covariate, and $\int_0^1 x_i(s) b^{(0)}(s) ds$ is the offset term. 
This approach is referred to as RAPLS-random hereafter.
}
% The RKHS approach was then implemented based on the algorithm described in Section 2 of \cite{yuan2010}, and the generalized cross-validation (GCV) statistic was used to select the optimal tuning parameter. 
% The NIAPLS was implemented using the R function {\tt fregre.pls()} in the R package {\tt fda.usc}, and the optimal number of FPLS components was also selected by GCV. The proposed RAPLS was implemented in the same way as that in Setting I. 

% We considered $n = 100$ and $200$, and the results are displayed in Fig. \ref{low}. 
% For estimation accuracy,  RAPLS outperforms RKHS and NIAPLS under all settings. For prediction accuracy,  RKHS performs similarly to the proposed RAPLS, both of which outperform  NIAPLS. 
Like Setting I, we generated 500 independent data sets for $n = 100, 200$ and $500$, and for each setting, we reported results for RAPLS, RAPLS-random, FPCR, and plsRglm. 
%while we reported $\mbox{MSE}(\alpha)$ for RAPLS, RAPLS-random-cal, FPCR, and plsRglm
% Results TBA ---
The results for $\mbox{MSE} (b)$  are displayed in Table. \ref{poisson}. 
%-----------------------------------------------
\begin{table}[h]
\centering
\begin{tabular}{cccccc}
\hline
$n$ & Scenario & RAPLS-random & RAPLS & FPCR & plsRglm \\
\hline
100 & I & 1.369 (0.433) & 1.371 (0.442) & 2.235 (0.891) & 4.434 (1.55)\\
200 & I & 0.379 (0.189) & 0.373 (0.179) & 0.447 (0.358) & 3.250 (1.04)\\
500 & I & 0.098 (0.037) & 0.096 (0.036) & 0.167 (0.189) & 2.233 (0.631)\\
100 & II & 1.279 (0.409) & 1.275 (0.402) & 2.603 (1.19) & 2.028 (0.613)\\
200 & II & 0.303 (0.103) & 0.298 (0.1) & 0.726 (0.601) & 0.786 (0.265)\\
500 & II & 0.094 (0.027) & 0.089 (0.027) & 0.396 (0.51) & 0.433 (0.136)\\
\hline
\end{tabular}
\caption{Simulation results for the FPM: $\mbox{MSE}(b)$ and its standard error over 500 independent data sets for $n = 100, 200, 500$: Scenario I corresponds to $b^*(s) = \frac{2}{3}\sum_{k=1}^{25} (-1)^k \phi_k(s)$, while Scenario II corresponds to $b^*(s) = \frac{2}{3}\sum_{k=26}^{50} (-1)^k \phi_k(s)$.}
\label{poisson}
\end{table}
%--------------------------------------------------
%%%%% Try different initial values ---
Like Table \ref{linear}, plsRglm show the worst performance in all cases except for Scenario II with $n = 100$, and 
RAPLS and RAPLS-random outperform FPCR and plsRglm in all settings. 
FPCR performs better in Scenario I than Scenario II, because $b^*(\cdot)$ in Scenario II
aligns with tail eigenfunctions, which require large sample sizes to estimate. 
%again demonstrating FPCR's limitation in estimating tail eigenfunctions. 
% In Scenario I, unlike Table \ref{linear} where RAPLS also dominates FPCR, FPCR shows lower MSE than RAPLS when the sample size is relatively large ($n = 200, 500$). 
RAPLS and RAPLS-random show similar results, demonstrating that the proposed RAPLS algorithm is robust to initial values. 
Results for $\mbox{MSE}(\alpha)$ show very similar patterns: both RAPLS methods outperform FPCR and plsRglm, while RAPLS and RAPLS-random show similar results. 
The estimation accuracy for the FPM is generally lower for all methods compared to the PFLM, likely due to the non-linear nature of the Poisson model.

\begin{table}[h]
\centering
\begin{tabular}{cccccc}
\hline
$n$ & Scenario & RAPLS-random & RAPLS & FPCR & plsRglm\\
\hline
100 & I & 0.227 (0.478) & 0.231 (0.482) & 0.368 (0.707) & 0.535 (0.265)\\
200 & I & 0.175 (0.369) & 0.172 (0.366) & 0.276 (0.576) & 0.372 (0.357)\\
500 & I & 0.037 (0.079) & 0.034 (0.075) & 0.053 (0.103) & 0.2 (0.283)\\
100 & II & 0.229 (0.478) & 0.232 (0.482) & 0.368 (0.708) & 0.544 (0.256)\\
200 & II & 0.171 (0.370) & 0.168 (0.367) & 0.276 (0.576) & 0.325 (0.362)\\
%\addlinespace
500 & II & 0.035 (0.073) & 0.033 (0.069) & 0.052 (0.095) & 0.2 (0.282) \\
\hline
\end{tabular}
\caption{Simulation results for the FPM: $\mbox{MSE}(b)$ over 500 independent data sets for $n = 100, 200, 500$: Scenario I corresponds to $b^*(s) = \frac{2}{3}\sum_{k=1}^{25} (-1)^k \phi_k(s)$, while Scenario II corresponds to $b^*(s) = \frac{2}{3}\sum_{k=26}^{50} (-1)^k \phi_k(s)$.}
\label{poisson}
\end{table}  
%%% Do this today -----
%We hypothesized that FPCR outperforms RAPLS in Scenario I because the initial values for RAPLS are sub-optimal. 

%in the supplementary file.

%----------------------------------------------------------

\section{Real data analysis} \label{sec:real}
We applied the proposed RAPLS algorithm to a sample of patients from the Alzheimer's Disease Neuroimaging Initiative (ADNI) study to analyze the progression of AD based on brain images. 
ADNI currently has  4 phases: ADNI1, ADNI-GO, ADNI2, and ADNI3, and its primary goal is to test whether serial magnetic resonance imaging (MRI), positron emission tomography (PET), and neuropsychological assessments can be used to measure the progression of AD. Participants were assessed at multiple visits. At each visit, various clinical measures, brain images, and neuropsychological assessments were collected. Detailed information about ADNI can be found at the official website \url{http://www.adni-info.org}. 

{
%\color{blue}
The stage of late mild cognitive impairment (LMCI) is considered a critical transitional stage between the normal stage and AD. However, it is unclear what brain regions drive the transition from LMCI to AD patients.
%In this application, we aim to 
%make a diagnosis of AD among individuals diagnosed as LMCI and to 
%identify important brain regions that may drive the AD progression based on the baseline PET images. 
PET neuroimaging directly measures the regional use of glucose with a lower glucose metabolic rate indicating less intensive neuronal activity, which has been proven as an important alternative to MRI images for AD diagnosis. 
We selected 302 subjects from ADNI1 without missing data in the covariates of interest. Among the 302 subjects, 95  subjects were diagnosed with AD before the study's completion and the remaining 207 individuals were diagnosed with LMCI; 107 were female, while 195 were male. The majority of the subjects were right-handed (283) compared to only 19 left-handed people. The subjects' ages span from 55 to 89 with the median being 75. 
%The demographic information of the selected subjects can be found in Section 5 of the supplementary document. 
}

We considered the following functional logistic model:
\begin{align}\label{real:adas}
\text{logit}(\mbox{pr}\{y_i = 1\mid {\bf z}_i, x_i(s)\}) = \alpha + {\bf z}_i^T \ve \beta + \int_{\mathcal{S}} x_i(s)b(s)ds ~~~\mbox{for~}  i = 1, \ldots, 236, 
\end{align}
where $\text{logit}(x) = \log(x) - \log(1-x)$, 
$y_i = 1$ indicates AD, and $y_i = 0$ indicates MCI. 
The covariates ${\bf z}_i$ includes  gender (1=Male; 0=Female), handedness (1=Right; 0=Left),
 and age. The functional predictor $x_i(s)$ is the PET imaging data measured on $160\times 160 \times 96$ voxels. The PET images underwent four preprocessing steps, which are introduced in detail in the supplementary document. We also removed the regions outside the skull, and around {900,000 voxels} remained. 
% \begin{figure}[t]
% \centering
% \includegraphics[width = 0.7\textwidth,height = 0.4\textheight]{ROC_ADNI1-eps-converted-to.pdf}
% \caption{ADNI results: ROC curves for FPCA, RAPLS, and RM.}\label{roc}
% \end{figure}
%----------------------------------------------------------------------------------

\begin{figure}[t]
\centering
\includegraphics[width =0.55 \textwidth,height = 0.3\textheight]{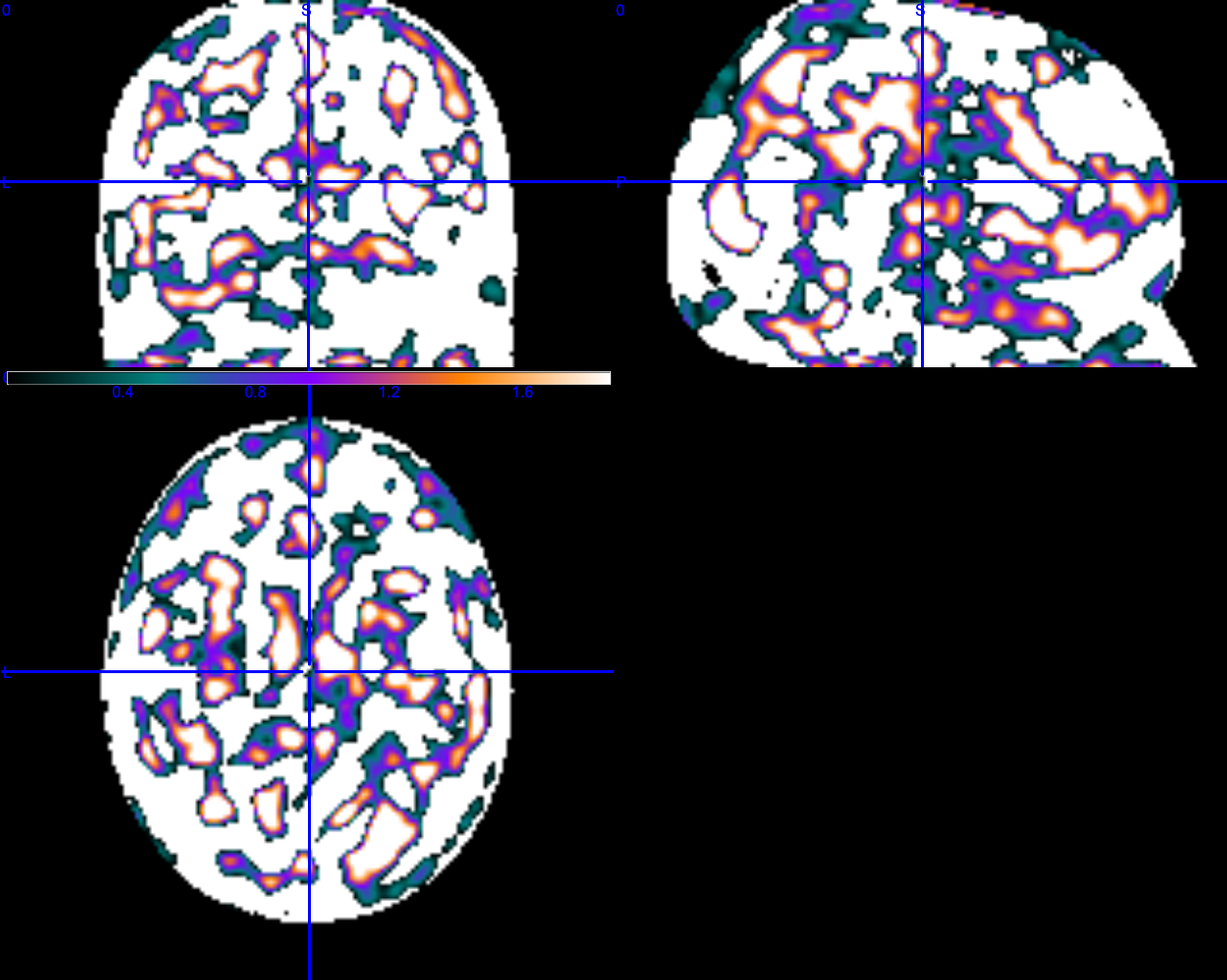}
\vspace{-1cm}
\caption{ADNI results:  the negative regions of estimated $b(s)$ using the entire cohort: RAPLS with 10 basis functions.  From left to right, each coefficient image is displayed with 3 slices in its transverse, coronal, and sagittal view located at \{80, 80, 48\}, respectively. }\label{fbeta:score:pca}
\end{figure}

{
%\color{blue}
Our primary analysis aims to identify brain regions that drive the AD progression from LMCI.
We estimated  $b(\cdot)$ in \eqref{real:adas} using the proposed RAPLS. 
%in terms of the estimation of $b(\cdot)$ in \eqref{real:adas}. 
%PlsRglm was not performed here because it was not able to handle images with 900,000 voxels. 
 %We estimated $b(s)$ in the model (\ref{real:adas}) by using all 236 subjects with both FPCA and RAPLS. 
The optimal number of RAPLS basis components, determined by AIC, was 11. 
%by (a) RAPLS with 3 basis functions and (b) FPCA with 14 basis functions. 
Fig. \ref{fbeta:score:pca}  presents three selected slices of the negative regions of both estimates of $b(s)$, where decreased glucose metabolism is associated with the progression to AD. % From the sagittal view, it can be seen that decreased glucose metabolism in the frontal, parietal, and occipital of the cerebrum are identified as important risk factors for the progression of AD. These areas are key to speech, concentration, language, vision interpretation, and memory, and thus decreased neuronal activities in these areas are likely to be indicators of AD progression. From the coronal view, we see that decreased neuronal activity in the hippocampus is also likely to be associated with the progression to AD. This has been demonstrated in several existing studies. 
This finding aligns with current biological understanding. Reduced glucose metabolism, is a well-documented feature in Alzheimer's Disease, particularly in regions such as the parietal lobe, temporal lobe, and posterior cingulate cortex \citep{sanabria2013glucose}. These areas are often implicated in cognitive functions like memory, attention, and spatial orientation, which are commonly affected in AD patients \citep{wilson2012natural}. 
%Thus, the observed relationship between lower glucose levels and increased AD risk in the highlighted regions from the figure is biologically plausible and consistent with existing research on metabolic impairments in Alzheimer's Disease. 
While the possibility of noise in the data should be acknowledged, the findings provide meaningful insights into the neurodegenerative processes in AD.
%For example, \cite{Chenbrainhippo2010} reported that hippocampal volume was associated with delayed verbal memory, which was an important predictor for determining whether a subject converted from MCI to AD or not \citep{Gomar2011}.  
}

%-----------------------------------------------------------------------------------
{
%\color{blue}
As a secondary analysis, we compared the predictive performance of RAPLS using a leave-one-out cross-validation (LOOCV) procedure to minimize variation in model evaluation. In addition, we included existing methods in the comparison, such as FPCR, plsRglm, linear discriminant analysis (LDA), and random forest (RF) models.
Since plsRglm, LDA, and RF require substantial memory and computational time to handle images with 900,000 voxels, we facilitated the comparison by subsampling the images at 1,000 equally spaced voxels across the entire domain. The dimension-reduced images were used for all downstream analyses. Additionally, because LDA suffers from singular covariance matrices in high-dimensional settings, we applied LDA to the top 50 principal components (PCs) of the reduced imaging data, which explained about 90\% of the total variation.
Based on the primary analysis, the optimal number of basis components was set to 11 for RAPLS. For FPCR and plsRglm, 8 and 14 basis components were selected based on AIC, respectively. The RF models were fitted using the R package {\tt randomForest} \citep{liaw2002classification}.
%While high-dimensional LDA tools have been recently developed \citep{tony2019high}, they require the calculation of a 900,000 $\times$ 900,000 (inverse) covariance matrix, which requires too much computational resources, if not infeasible.
%without relying on dimension reduction was implemented \cite{}, referred to high-LDA. 
%This approach 
%We first compared FPCA with the proposed RAPLS algorithm based on the leave-one-out cross-validation; that is, for $i = 1, \ldots, 236$,
%we fitted model (\ref{real:adas}) using all samples except for the $i$-th sample and a prediction of the probability of being classified as AD is made for the $i$-th sample.  
% In addition, to demonstrate the predictive value of the PET images in addition to the demographic covariates, we also considered a reduced model (RM) without imaging predictors as a benchmark:
% \begin{align*}
% \text{logit}(\mbox{pr}(y_i = 1 \mid {\bf z}_i, x_i(s))) = \alpha + {\bf z}_i^T \ve \beta ~~~\mbox{for~}  i = 1, \ldots, 236.
% \end{align*}
%Based on RM, the probability of having AD was predicted for each sample using the aforementioned leave-one-out cross-validation procedure.
%We created the ROC curves based on the predictions of all methods in Fig. \ref{roc}. Inspecting Fig. \ref{roc} shows the predictive value of the PET images regarding the AD progression, as both FPCA and RAPLS yield much higher classification accuracy than RM. Also, the area under the curve (AUC) for RAPLS is 0.71 compared to 0.67 for FPCA, leading to a 6\% improvement in the classification accuracy. 
The prediction accuracy for RAPLS, FPCR, plsRglm, LDA, and RF was 68.5\%, 66.6\%, 64.2\%, 63.9\%, and 66.9\%, respectively. Notably, RAPLS achieved the highest prediction accuracy among the methods.
Like the simulation results, RAPLS outperformed plsRglm, another method based on partial least squares. 
%highlighting the superior predictive performance of our approach. 
This suggests that the IRLS-based construction of PLS basis functions in RAPLS is more suitable for nonlinear models compared to the GLM-based construction used in plsRglm in terms of prediction. 
%However, when we delved into the performance of RAPLS and FPCR, we realized that FPCR classified every patient as LMCI, leading to the classification accuracy of $207/302 \approx 68.5\%$. 
%RAPLS classified 95 patients as AD with 25 of them being correctly 
%However, as it is well known that the difference between LMCI and early AD tends to be vague, all these prediction accuracies are not very high. 

We should acknowledge that the distinction between LMCI and early-stage AD is often subtle and difficult to define. Clinically, the transition from LMCI to early AD is marked by a gradual progression of symptoms, with considerable overlap in cognitive decline, memory impairment, and other neurodegenerative indicators. As a result, the biological and clinical boundaries between these two groups are inherently blurred. This overlap makes it challenging to develop models that can consistently and accurately predict the correct classification. Consequently, as observed in prior studies, prediction accuracies tend to be low, which is consistent with existing studies with even larger sample sizes \citep{nozadi2018classification}. 
}
% back to edit

%%%% we need to do this --------
%The proposed calibration procedure was not implemented in this data application because two binary covariates (gender and handedness) cannot be well characterized using the linear model (\ref{add:022822:1}). It would be a fruitful future project to extend the calibration procedure to handle non-continuous scalar covariates. 

%--------------------------------------------------------------
\section{Discussions}

{
%\color{blue}

This paper introduces a residual-based alternative partial least squares (RAPLS) method for parameter estimation in a class of generalized functional linear models (GFLM). The key idea behind RAPLS is the integration of the iteratively reweighted least squares (IRLS) and the alternative partial least squares (APLS), enabling accurate approximation of nonlinear functional models through a sequence of functional linear models. RAPLS demonstrates clear advantages in both estimation and prediction, particularly when the underlying function parameter is closely aligned with the tail eigenfunctions.

Identifiability is a general challenge in functional regression models.
For example, in a simple functional linear model, $y = \alpha + \int x(s)b(s) ds + \epsilon$, the function $b(\cdot)$ is not identifiable if $x(\cdot)$ lies within a low-dimensional space $\mathcal{S}$, but $b(\cdot)$ does not. Specifically, any part of $b(\cdot)$ that lies outside of $\mathcal{S}$ will remain unidentifiable. This challenge persists when dimensionality reduction techniques are applied to estimate $b(s)$. For instance, when using the top $p$ RAPLS basis functions, the estimated parameter is $b_p(\cdot)$, the projection of $b(\cdot)$ onto the space spanned by the top $p$ RAPLS basis functions. In this case, the difference $b - b_p$ remains unidentifiable. This issue is particularly concerning if $\|b - b_p\|$ does not converge to 0 as $p \rightarrow \infty$. Fortunately, RAPLS overcomes this issue, as Lemma \ref{lemma2} guarantees that $\|b - b_p\|$ diminishes as more basis functions are used. A similar result holds for FPCR when eigenfunctions are employed as basis functions, provided that the covariance kernel of the functional covariates is positive definite.

This work is motivated by brain imaging applications where the images are regularly spaced. However, we acknowledge that in many other applications, such as those involving longitudinal designs, functional data may be sparse or measured irregularly. While RAPLS is methodologically applicable to such data, we anticipate that some pre-smoothing will be necessary in these cases. Evaluating the performance of RAPLS on irregularly observed functional data will be left as a future research direction.

Given the connectivity structures of the brain, the functional parameter associated with brain images is typically assumed to be smooth across voxels. In other contexts, achieving sparsity in the estimated coefficient function may be desirable for greater interpretability. To induce sparsity in the coefficient function, one potential approach is to introduce sparsity into the RAPLS basis functions. In non-functional partial least squares (PLS), sparsity can be achieved by adding constraints to the iterative algorithms. For example, consider the linear model $Y = X\beta + \epsilon$. The first ``sparse" PLS basis function can be constructed by solving:
\[
\max {\bf w}^\intercal X^\intercal Y Y^\intercal X {\bf w}, ~~~ \mbox{subject to  } \|{\bf w}\|_2 = 1, \|{\bf w}\|_1 < \lambda,
\]
where $\lambda$ controls the level of sparsity. Extending this idea to RAPLS presents an exciting opportunity for future research.

}

%---------------------------------------------------------------
\section*{Supplementary Materials}

Online supplementary material includes additional simulations and theoretical results, proofs of the main theorems, and supporting information for the real data application. 
\par
%%%%%%%%%%%%%%%%%%%%%%%%%%%%%%%%%%%%%%%%%%%%%%%%%%%%%%%%%%%%%%%%%%%%%%%%%%%%%%%%%%%%%%%%%%%%%%%%%%%%%%%%%%%%%%%%%%%%%%%%%%%%
\section*{Acknowledgements}

Data used in the preparation of this article were obtained from the Alzheimer's Disease Neuroimaging Initiative (ADNI) database. As such, the investigators within the ADNI contributed to the design and implementation of ADNI and/or provided data but did not participate in the analysis or writing of this report. A complete listing of ADNI investigators can be found at 
\url{http://adni.loni.usc.edu/wpcontent/uploads/how to apply/ADNI Acknowledgement List.pdf}. 
Dr. Zhu was partially supported by the National Institutes of Health (NIH) grants 1R01AR082684,  1OT2OD038045-01, and   the National Institute on Aging (NIA) of the National Institutes of Health (NIH) grants  U01AG079847, 1R01AG085581, RF1AG082938, and R01AR082684.
\par

%%%%%%%%%%%%%%%%%%%%%%%%%%%%%%%%%%%%%%%%%%%%%%%%%%%%%%%%%%%%%%%%%%%%%%%%%%%%%%%%%%%%%%%%%%

\bibhang=1.7pc
\bibsep=2pt
\fontsize{9}{14pt plus.8pt minus .6pt}\selectfont
\renewcommand\bibname{\large \bf References}
%\begin{thebibliography}{11}
\expandafter\ifx\csname
natexlab\endcsname\relax\def\natexlab#1{#1}\fi
\expandafter\ifx\csname url\endcsname\relax
  \def\url#1{\texttt{#1}}\fi
\expandafter\ifx\csname urlprefix\endcsname\relax\def\urlprefix{URL}\fi

% use bibfile 
 \bibliographystyle{chicago}      % Chicago style, author-year citations
 \bibliography{plsbib,ZhuShenR01July11,finalsurvbib, paper-ref}   % name your BibTeX data base

%%%%%%%%%%%%%%%%%%%%%%%%%%%%%%%%%%%%%%%%%%%%%%%%%%%%%%%%%%%%%%%%%%%%%%%%%%%%%%%%%%%%%%%%%%%%%%%%%%%%%%%%%%%%%%%%%%%%%%%%%%%%
\vskip .65cm
\noindent
Yue Wang
\vskip 2pt
\noindent
E-mail: yue.2.wang@cuanschutz.edu
\vskip 2pt

\noindent
Xiao Wang
\vskip 2pt
\noindent
E-mail: wangxiao@purdue.edu

\noindent
Joseph G. Ibrahim
\vskip 2pt
\noindent
E-mail: ibrahim@bios.unc.edu

\noindent
Hongtu Zhu
\vskip 2pt
\noindent
E-mail: htzhu@email.unc.edu

% \vskip .3cm
%\centerline{(Received ???? 20??; accepted ???? 20??)}\par
\end{document}